# A crayfish-optimized wavelet filter and its application to fault diagnosis

Sumika Chauhan[1], Govind Vashishtha[1*], Radoslaw Zimroz[1] and Rajesh Kumar[2]
[1]Faculty of Geoengineering, Mining and Geology, Wroclaw University of Science and Technology, Na Grobli 15, 50-421 Wroclaw, Poland
[2]Precision Metrology Laboratory, Department of Mechanical Engineering, Sant Longowal Institute of Engineering and Technology, Longowal 148 106, India
* Corresponding author, Email address: govindyudivashishtha@gmail.com
ORCID iD: 0000-0002-5160-9647

**Abstract:** Industrial machine fault diagnosis ensures the reliability and functionality of the system, but identifying informative frequency bands in vibration signals can be challenging due to low signal-to-noise ratio (SNR), background noise, and random interferences. The wavelet filter is commonly used for this purpose, but its parameters are crucial for locating the informative frequency band to extract repetitive transients. This study utilizes a crayfish optimization algorithm (COA) to optimize the wavelet filter adaptively for extracting fault characteristics. COA uses correlated kurtosis (CK) as a fitness function while addressing issues related to inaccurate CK period through an updation process. The proposed methodology is applied to different industrial cases and compared with existing methods, demonstrating its superiority in extracting informative frequencies.

## 1. Introduction

Bearings and gears play a vital role in the operation of rotating machinery, serving as crucial elements in a wide range of industrial uses. Their failure can lead to significant operational disruptions or catastrophic events. Consequently, early fault detection is crucial and has garnered extensive attention [1][2]. Vibration analysis is a commonly employed method for identifying specific faults in these parts through the detailed fault data present in the vibration signals obtained from bearings and gears [3][4]. When a defect arises in these parts, it results in regular impacts that stimulate high-frequency resonance within the bearing system. These impacts create pulses with frequencies referred to as characteristic frequencies, which are distinctively defined by the rotational speed, position of the fault, and geometric properties of the rotating components. Therefore, the identification of these characteristic frequencies is a key method for detecting faults in rotating components. The main goal of fault detection is to identify these characteristic frequencies to diagnose the condition of the rotating components accurately [5][6].

However, detecting defects continues to be a challenging issue due to the demanding working conditions and intricate structures of mechanical systems. The challenges in diagnosing these issues stem from several factors: Defect impulses resulting from various faults may overlap in the time domain, making it difficult to differentiate between them [7]. Moreover, these impulses can be masked by significant background noise, further complicating the fault detection process [8][9]. Different faults may trigger either similar or distinct resonance frequencies, adding complexity to the diagnosis as it becomes more difficult to accurately attribute specific resonance frequencies to particular faults [10]. Overcoming these obstacles is crucial for enhancing the reliability and precision of these rotating components.

Several methods have been researched to effectively identify faults in rotating components. One of these approaches is spectral kurtosis (SK), which was introduced by Antoni and Randall to determine the presence of fault information within a frequency band, aiding in the selection of an optimal frequency band for demodulation [11][12][13]. Lei et al. [14] suggested an enhanced Kurtogram method using a wavelet packet transform filter bank to overcome limitations of the original Kurtogram technique. Building on this work, Wang et al. [15] enhanced the Kurtogram by devising a method to compute kurtosis utilizing the power spectrum of envelope signals derived from wavelet transform. Additionally, Tian et al. [16] introduced a new technique for extracting fault features, which integrates the Gini indices based on maximum envelope spectrum power function with empirical wavelet transform. This method represents a development from the existing Fast Kurtogram approach. Peng et al. [17] proposed the adaptive reweighted-kurtogram method to effectively extract diagnostic information while dealing with strong impulse noise. Guo et al.'s [18] paper introduced a new approach called "differgram" for fault diagnosis in rotating machinery, aiming to overcome limitations of existing techniques such as fast kurtogram and its variants. Liu et al. [19] presented the DTMSgram method to address issues where fault signatures are often masked by complex background noise from wheel-rail excitation and transmission paths, intending to enhance fault feature components in vibration signals and improve demodulation frequency band selection accuracy. Despite the effectiveness of Kurtogram-based methods in detecting faults, they have inherent deficiencies, such as challenges in identifying and selecting multiple frequency bands for demodulation and resonance frequency band splitting caused by the fixed tiling pattern of the Kurtogram.

Wiggins [20] introduced the minimum entropy deconvolution, which is a commonly utilized technique for fault detection. The core idea behind MED is to find an FIR filter that reduces the entropy of the filtered signal. Vashishtha et al. [21] proposed a method for detecting bearing defects in Francis turbines using MED filters applied to acoustic signals. This approach employed Aquila optimizer to dynamically determine the optimal filter length by considering autocorrelation energy as its fitness function, aiming to isolate impulses in weak signals. Xie et al. [22] introduced a novel approach for early fault detection in rolling bearings, which focuses on adaptive minimum noise amplitude deconvolution. This method aims to overcome the difficulty of identifying subtle fault characteristics that may be hidden by background noise. Although MED has been successfully used for fault detection, it relies on kurtosis as an indicator and may be sensitive to random impulses, potentially compromising accuracy and reliability in fault detection tasks. McDonald et al. [23] presented maximum correlated kurtosis deconvolution (MCKD) to alleviate some limitations associated with MED. This approach is designed to maximize the CK of the resulting signal, highlighting both the suddenness and regularity of defective impulses. However, achieving optimal performance with MCKD necessitates accurate input parameters such as fault period, filter length, maximum iterations, and shift order. While methods like MED and MCKD represent notable progress in fault detection, their effectiveness may be constrained by the requirement for precise input parameters and potential suppression of specific frequency ranges. Additional research and fine-tuning are needed to address these obstacles and enhance the dependability of these techniques for identifying faults in rolling element bearings.

In addition to these approaches, various techniques in signal processing have been effectively utilized for detecting faults in industrial components. An advancement known as the self-iterated extracting wavelet transform was introduced by Li et al. [24], providing a notable improvement in time-frequency analysis, particularly suitable for signals exhibiting strong amplitude and frequency modulation. Wang et al. [3] combined the improved deep residual network with wavelet transform to address the challenge of imbalanced healthy state data versus faulty state data. Cui et al. [25] introduced an adaptive approach to kernel sparse representation classification, which overcomes challenges associated with time-shift properties. This method utilizes data fusion across multiple domains, leading to improved dictionary learning for intelligent bearing fault diagnosis and superior feature representation as well as classification performance. Li et al. [26] introduced a new approach to fault diagnosis in rotating machinery, utilizing multi-scale feature extraction and domain adaptation techniques. This method effectively captures both local and global features from vibration signals while adapting to differences between domains, resulting in improved diagnostic accuracy and robustness under challenging operating conditions. Meanwhile, Liu et al. [27] developed the random spectral similar component decomposition to address issues related to sampling processes and mode mixing encountered with various decomposition techniques. Additionally, Chauhan et al. [28] proposed a novel scheme for detecting bearing defects using the Single-Valued Neutrosophic Cross-Entropy combined with an adaptive feature mode decomposition method. Finally, Huo et al. [29] introduced a method for adaptive time-frequency decomposition that captures specific time-frequency characteristics using ridge extraction methods.

To address the limitations of the above methods, a novel scheme of fault diagnosis is proposed based on Morlet wavelet filter and COA for selecting the optimal frequency band for demodulation. The COA is used to optimize the center frequency ($f_c$) and bandwidth ($\sigma$) taking CK as the fitness function. The proposed method incorporates an update process to ensure that the period gradually approaches the real value, enhancing the accuracy of the correlated kurtosis. This iterative process ensures that the optimal frequency band is selected accurately, improving the detection of faults.

The rest of the paper is structured as follows. Section 2 provides an overview of the fundamental concepts related to wavelets COA and CK. The approach put forward by this study is outlined in Section 3, while Section 4 elaborates on the process of applying this approach to industrial machinery. A comparison between the proposed method and established techniques is presented in Section 5. Finally, conclusions are drawn in Section 6.

## 2. Premilinaries

### 2.1. Theory of wavelets

Wavelets are a powerful mathematical tool used for analyzing and representing signals and functions at multiple levels of detail. The wavelet is expressed as scaling and shifting of the mother wavelet $\psi(t)$

$$\psi_{(a,b)}(t) = |a|^{-1/2}\psi\left(\frac{t-b}{a}\right) \tag{1}$$

The scale parameter and time parameter are denoted as $a$ and $b$, respectively. The CWT of signal $x(t) \in L^2(R)$ is expressed as the inner product in the Hilbert space.

$$W_b(a) = |a|^{-1/2}\int_{-\infty}^{\infty} x(t)\psi^*\left(\frac{t-b}{a}\right)dt \tag{2}$$

where asterisk is a complex conjugate.

Using the convolution theorem and the scaling property of the Fourier transform (FT), it is possible to express Eq. (2) in an alternative form.

$$W_b(a) = \frac{1}{\sqrt{a}}F^{-1}\{X(f)\psi^*(af)\} \tag{3}$$

where $F^{-1}$ denotes the inverse Fourier transform (IFT), and $X(f)$ and $\psi(f)$ represent the FT of $x(t)$ and $\psi(t)$, respectively. Eq. (3) demonstrates that the CWT of signal $x(t)$. The Morlet wavelet is defined as [30].

$$\psi(t) = ce^{-\sigma^2 t^2} e^{j2\pi f_c t} \tag{4}$$

The parameter $c$ is a positive parameter and can be obtained through Eq. (5)

$$c = \sigma/\sqrt{\pi} \tag{5}$$

By combining Eqs. (4) and (5), we can express the FT of the Morlet wavelet as follows.

$$\psi(f) = e^{-(\pi^2/\sigma^2)(f-f_c)^2} \tag{6}$$

where $f_c$ represents the center frequency and $\sigma$ signifies the bandwidth. Consequently, the Morlet wavelet's associated frequency band, with a Gaussian shape in frequency domain, is confined within the interval $[f_c - \sigma/2, f_c + \sigma/2]$. The wavelet filter can be represented by the signal passing through it as

$$WT(f_c, \sigma) = F^{-1}\{X(f)\psi^*(f)\} \tag{7}$$

## 2.2. Crayfish optimization algorithm (COA)

The COA is a nature-inspired optimization algorithm based on the social behaviour of crayfish [31]. This algorithm mimics the social interactions and foraging behaviour of crayfish to solve optimization problems. The COA begins with an initial population of candidate solutions, which are then improved iteratively through a series of steps inspired by crayfish behaviour. These steps include exploration, exploitation, and communication among the individuals in the population. By simulating the social and foraging behaviours of crayfish, the COA aims to find the optimal solution to a given problem. In recent years, the COA has gained attention in the optimization community due to its effectiveness and potential for solving real-world problems. Overall, the COA offers a promising approach to solving optimization problems, and its unique inspiration from crayfish behaviour sets it apart from traditional optimization techniques. The procedure of COA is elaborated in the following subsections.

### 2.2.1. Initialization

The COA starts with random initialization to produce potential solutions $X$ with a specified population size $N$ and dimension $dim$. The location $X_{i,j}$ of individual $i$ in dimension $j^{th}$ is modeled in Eq. (8).

$$X_{i,j} = lb_j + (ub_j - lb_j) \times rand \tag{8}$$

Where $ub_j$ and $lb_j$ are upper and lower bounds of the $j^{th}$ dimension.

### 2.2.2. Defining the temperature and number of crayfish

The temperature plays a noteworthy role in the different stages of crayfish, as indicated in Eq. (9). When the temperature exceeds 30°C, the crayfish seeks cooler locations for its summer retreat. Within an optimal temperature range of 15°C to 30 °C, the crayfish initiates its foraging activities. The foraging behaviour can be represented by a normal distribution due to its temperature sensitivity, which is mathematically described in Eq. (10).

$$temp = rand \times 15 + 20 \tag{9}$$

$$p = C_1 \times \left( \frac{1}{\sqrt{2 \times \pi} \times \sigma} \times exp\left(-\frac{(temp - \mu)^2}{2\sigma^2}\right) \right) \tag{10}$$

where, $temp$ represents the temperature at which the crayfish is situated. The variable $\mu$ signifies the highest temperature experienced by the crayfish. Additionally, variables like "$\sigma$" and "$C_1$" manage the intake of crayfish at varying temperatures.

### 2.2.3. Summer resort stage

If temperature exceeds 30 °C, the crayfish retreats to its cave $X_{shade}$ for summer vacation, which is defined by Eq. (11).

$$X_{shade} = (X_G + X_L)/2 \tag{11}$$

The $X_G$ is best position achieved so far, $X_L$ is current population's location. The battle for cave occurs randomly. When $rand < 0.5$, there is no competition, crayfish directly take possession of cave as indicated in Eq. (12).

$$X_{i,j}^{t+1} = X_{i,j}^t + C_2 \times rand \times (X_{shade} - X_{i,j}^t) \tag{12}$$

Here, t is the current position, $t + 1$ is the next position. The parameter $C_2$ is the decreasing curve computed by Eq. (13).

$$C_2 = 2 - (t/T) \tag{13}$$

The value of $T$ is maximum number of iterations. This behaviour imitates the way crayfish approach the cave and mimics the process of finding an ideal solution.

### 2.2.4. Competition stage

The other crayfish are attracted to the same cave if $temp > 30$ and $rand \geq 0.5$ and as a result, they engage in fights to claim ownership of it, as demonstrated by Eq. (14).

$$X_{i,j}^{t+1} = X_{i,j}^t - X_{z,j}^t + X_{shade} \tag{14}$$

where, $z$ is the arbitrary crayfish and obtained by Eq. (15).

$$z = round(rand \times (N-1)) + 1 \tag{15}$$

### 2.2.5. Foraging stage

The crayfish begins foraging for food when temperature drops below $30°C$. Upon locating the food, the crayfish assesses its size using Eq. (16) and (17) to determine both the location and dimensions of the food item.

$$X_{food} = X_G \tag{16}$$

$$Q = C_3 \times (fitness_i / fitness_{food}) \tag{17}$$

The crayfish breaks down its food into smaller pieces using its claws, which imitates the process of seeking the most efficient solution.

### 2.3. Correlated Kurtosis (CK)

The researchers have used various criteria such as Shannon entropy, Gini index, kurtosis, $l_2/l_1$ norm to extract sensitive information. However, many of these measures are based on statistics and may not take into account the periodic nature of fault impulses in signals from rotating machinery. To overcome this limitation, CK has been utilised as a fitness function to optimize the morlet wavelet function which is a valuable metric used in fault detection, particularly for rotating machinery, because it considers the periodicity of defect impulses in the vibration signals. Unlike traditional measures such as Shannon entropy or simple kurtosis, which might miss the periodic nature of fault-related impulses. The CK provides a more robust means to detect these periodic impulses whose expression is given in Eq. (18)

$$CK_M(T_s) = \frac{\sum_{n=1}^{N}(\prod_{m=0}^{M} y_n - mT_s)^2}{(\sum_{n=1}^{N} y_n^2)^{M+1}} \tag{18}$$

where the shift order is represented by $M$ and $T_s$ denotes the number of sampling points during each fault period:

$$T_s = f_s * T \tag{19}$$

where $f_s$ and $T$ represent the sampling frequency and the fault period, respectively.

### 3. Proposed methodology

The selection of the optimal resonance frequency band plays a vital role in demodulation for extracting the fault characteristics. Researchers have put forward different methods including windowing, spectral kurtosis, kurtogram etc. In this work, the Morlet wavelet filter has been utilised to find the optimal frequency band as the shape of the Morlet wavelet precisely resembles the fault features in a noisy signal. Also, the Morlet wavelet filter has only two parameters: center frequency ($f_c$) and bandwidth ($\sigma$). This simplicity facilitates easy design and implementation. However, the Morlet wavelet function is a complex exponential function that is modulated by a Gaussian function. The $f_c$ and $\sigma$ of the Morlet wavelet significantly impact its time-frequency resolution. The $f_c$ determines the oscillatory cycle of the wavelet, with higher frequencies resulting in more oscillations within a given time window. On the other hand, the $\sigma$ affects the spread of the wavelet in the frequency domain, with broader $\sigma$ leading to a wider spread of frequencies that the wavelet can effectively capture. Adjusting the $f_c$ allows for the selection of specific frequency bands of interest, while modifying the bandwidth influences the trade-off between time and frequency resolution. A narrower $\sigma$ provides sharper frequency discrimination at the expense of temporal resolution, while a wider bandwidth offers better time localization at the cost of frequency precision. Therefore, understanding the influence of $f_c$ and $\sigma$ on the Morlet wavelet function is crucial for effectively analyzing time-frequency representations of signals. To address these issues, $f_c$ and $\sigma$ of the Morlet wavelet have been optimised by CAO using the correlated kurtosis as fitness function. The optimized wavelet filter is further utilised to find the optimal resonance frequency band. The procedure for the porposed methodology is elaborated through the following steps:

1. Input raw signal.
2. Initialize the parameters of the CAO.
3. Identify the most effective search agents by evaluating the fitness function through CK.
4. The unprocessed signal is passed through an optimal Morlet wavelet filter, selected to maximize the CK of the filtered signal in this specific iteration.
5. Calculate the square of the envelope spectrum of the filtered signal, which is then used to determine the characteristic frequencies.

The proposed approach is also illustrated by a flow diagram as shown in Fig. 1.

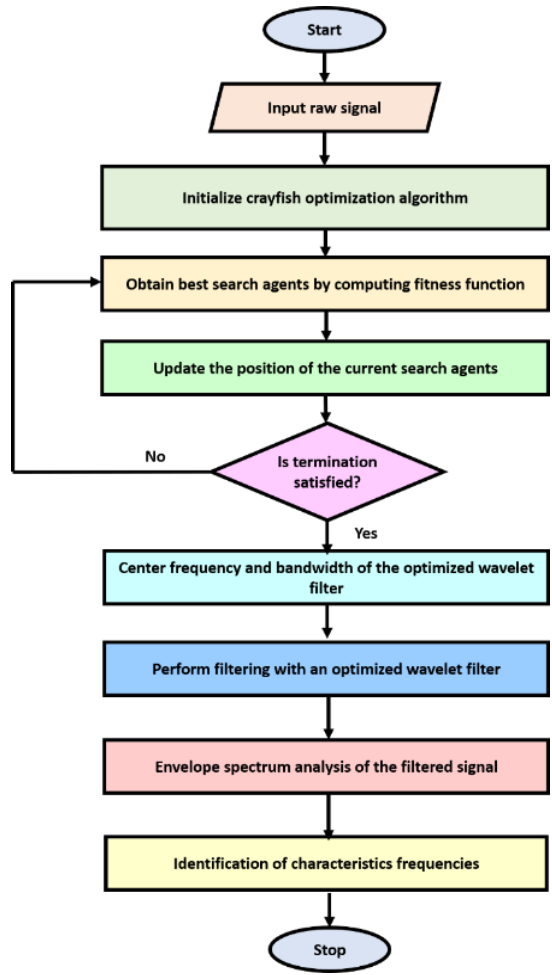

**Fig. 1** Proposed methodology's flowchart

## 4. Application of the proposed methodology on industrial machines

In mining industries, complex machinery with various rotating parts is commonly observed. An example of this is the conveyor driving station illustrated in Fig. 2, which incorporates a two-stage gearbox, an electric motor coupling, and a pulley operated by two 1000 kW drives. The pulley is positioned on a shaft supported by dual sets of bearings and connected to the gearbox through a coupling. The belt motion is facilitated by idlers while vibration signals from the supporting bearings have been recorded using an accelerometer mounted horizontally as depicted in Fig. 2 with a screw attachment. A sampling frequency of 19.2 kHz has been set, and the fault frequency has been identified at 12.6 Hz.

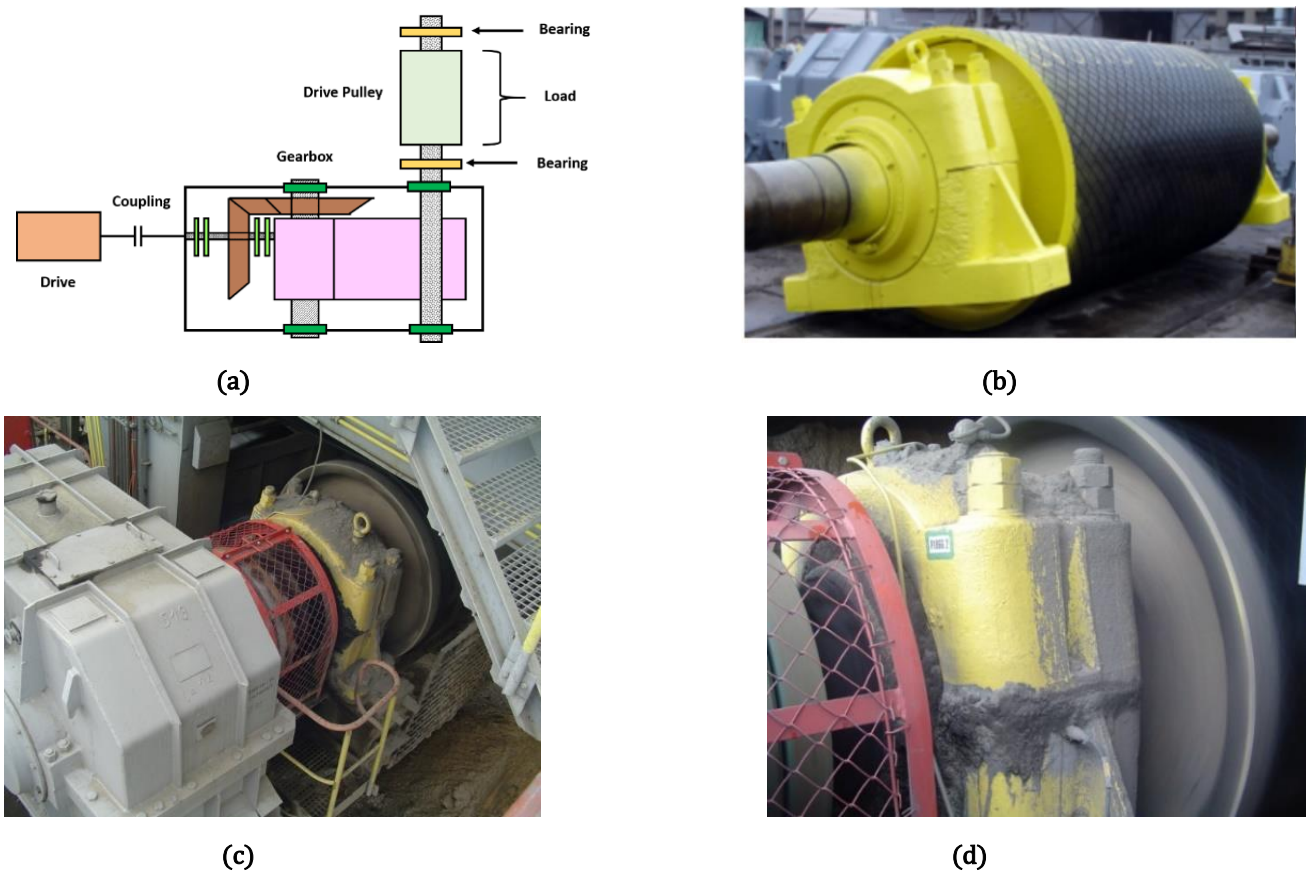

(a)　　　　　　　　　　　　　　　　(b)

(c)　　　　　　　　　　　　　　　　(d)

Fig. 2. A drive unit for belt conveyor

## 4.1. Vibration signal from bearing

The time and frequency domain waveforms of the faulty bearing in the belt conveyor system are displayed in Fig. 3. The significant noise and interferences from other components obscure the informative signals in the time-domain signal, making it challenging to identify fault features. Additionally, the corresponding envelope spectrum does not provide sufficient information to detect the characteristic frequency associated with bearing defects. However, the suggested approach demonstrates improved capability in extracting informative frequencies, as illustrated in Fig. 4.

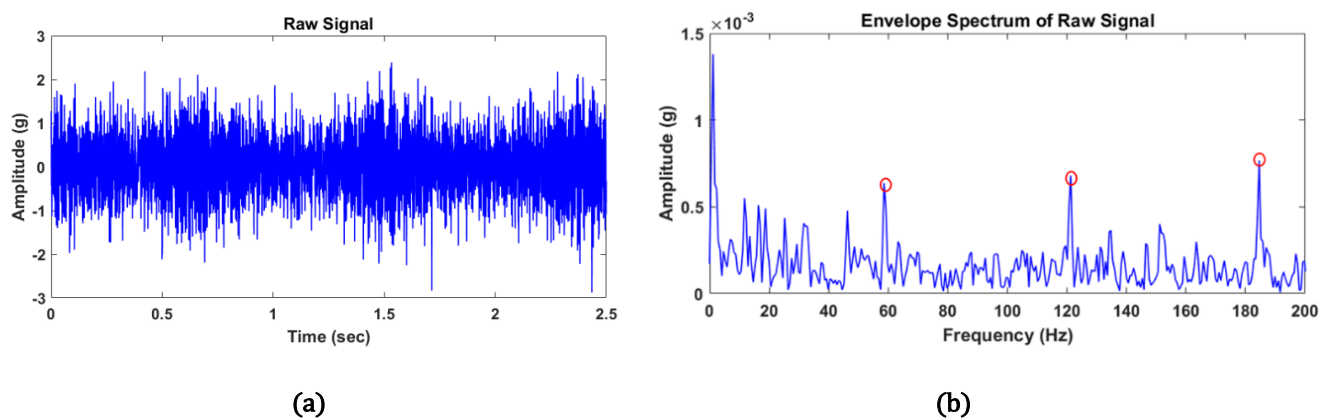

(a)　　　　　　　　　　　　　　　　(b)

Fig. 3. Vibration signal from defective bearing (a) Raw signal (b) Envelope spectrum of the raw signal

The signal processed using the proposed method is shown in Fig. 4, revealing periodic impulses in the time waveform (Fig. 4a). The kurtosis and SNR of the processed signal exhibit significant improvement. Additionally, the spectrum clearly displays the fault frequency of outer race defect (12.6 Hz) and its harmonics.

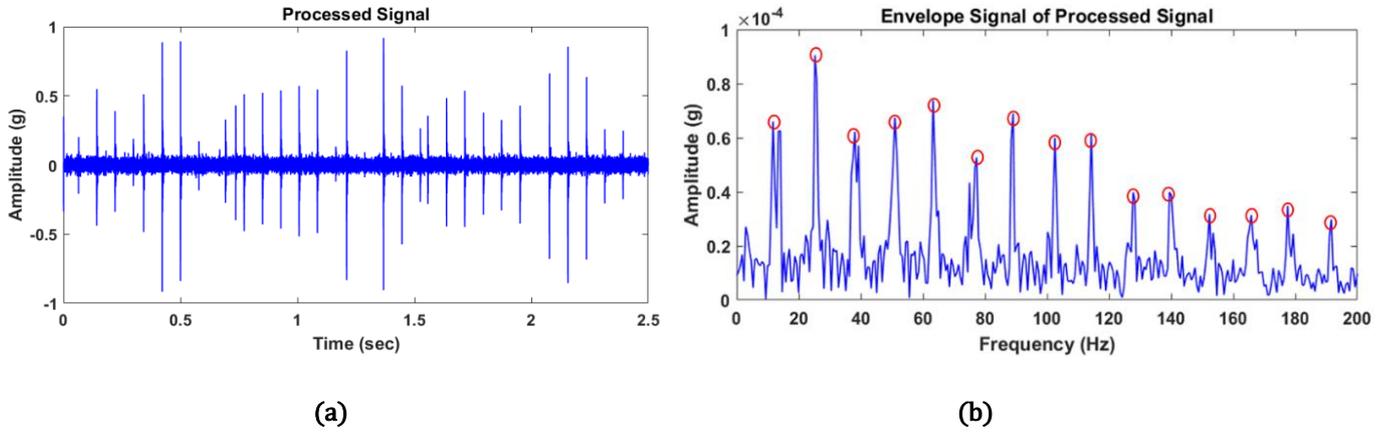

(a)          (b)

**Fig. 4.** Vibration signal from defective bearing (a) Processed signal (b) Envelope spectrum of the processed signal

### 4.2. Vibration signal from two-stage gearbox

A vibration signal is gathered from the two-stage gearbox, which transmits power from the engine to the driving pulley of the belt conveyor. The accelerometer is positioned on the gearbox as depicted in Fig. 5. A 2.5-second signal is captured at a sampling rate of 8192 Hz. The rpm of the input shaft remains constant, disregarding minor variations, and a fault is detected at a frequency of 4.1 Hz.

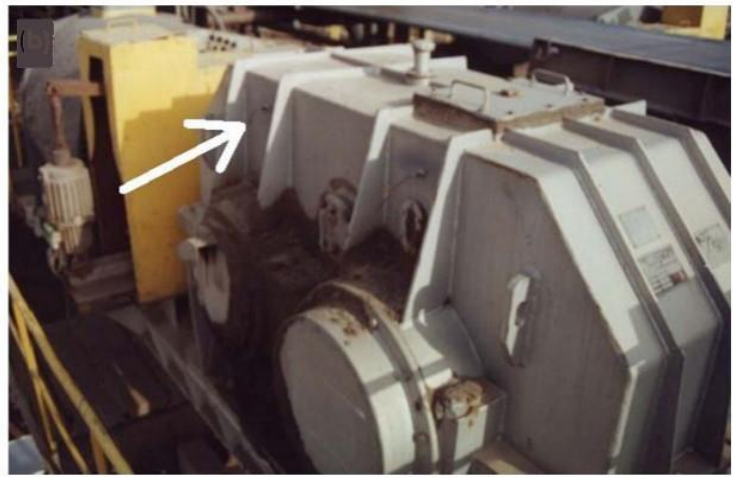

**Fig. 5.** Data acquisition from a two-stage gearbox

The time domain signal and its spectrum acquired from the gearbox are shown in Fig. 6. The kurtosis and SNR values of the original signal are notably low, which complicates the detection of periodic impulses within the signal. While some harmonics of the fault frequency can be identified in the envelope spectrum depicted in Fig. 6 (b), the distinctive fault frequency and its additional harmonics are fully obscured by numerous interferences from other components, impacting the accuracy of fault diagnosis.

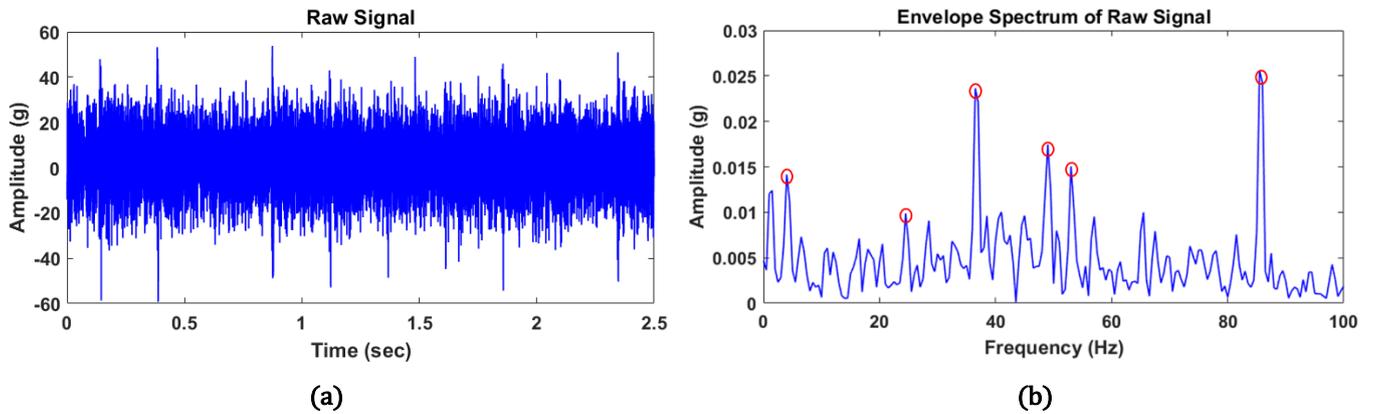

Fig. 6. Vibration signal from two-stage gearbox (a) Raw signal (b) Envelope spectrum of the raw signal

The proposed approach is used to analyze the vibration signal from the gearbox. Fig. 7 presents the time and frequency domain representation of the signal using this method. In Fig. 7 (a), periodic impulses are evident in the time domain signal, while in the spectrum, one can observe the characteristic fault frequency and its harmonics.

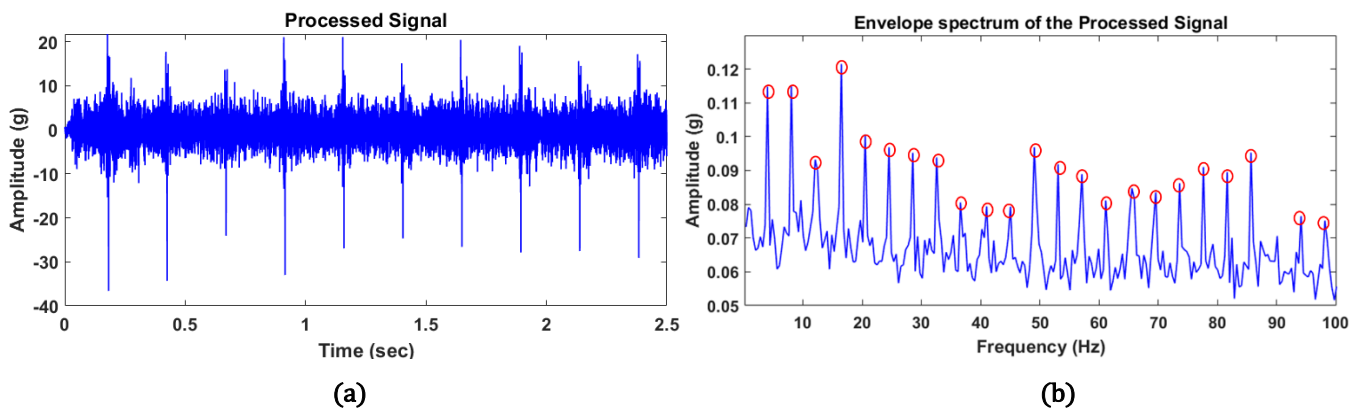

Fig. 7. Vibration signal from two-stage gearbox (a) Processed signal (b) Envelope spectrum of the processed signal

5. Comparison of the proposed method with

The suggested scheme is compared with existing methods to validate its effectiveness. The existing methods include fast kurtogram [12] and fast spectral coherence [32].

The fast kurtogram-based approach has also been employed for comparison. The signals from a faulty bearing and two-stage gearbox were examined using the fast kurtogram technique. Fig. 8 shows the time domain signal, while Fig. 9 presents its envelope spectrum for the defective bearing and two-stage gearbox, respectively. Regarding a defective bearing, the fast kurtogram is capable of identifying periodic impulses with significantly higher values of kurtosis and SNR compared to the raw signal. Nonetheless, our proposed method yielded superior outcomes compared to the kurtogram-based approach in this case. Furthermore, utilizing the kurtogram-based method facilitated easy identification of fault frequency at 12.6 Hz as well as its harmonics; however, it also revealed other frequencies beyond the fault frequency that could impact fault diagnosis results.

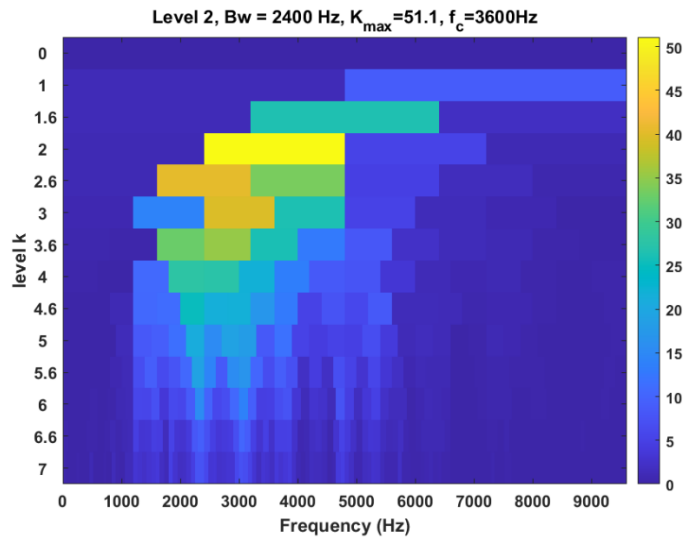

(a)

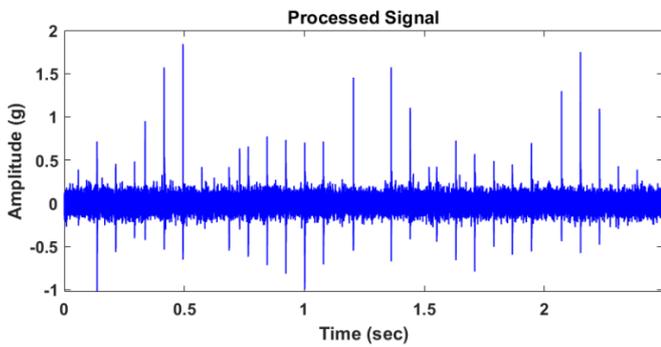

(b)

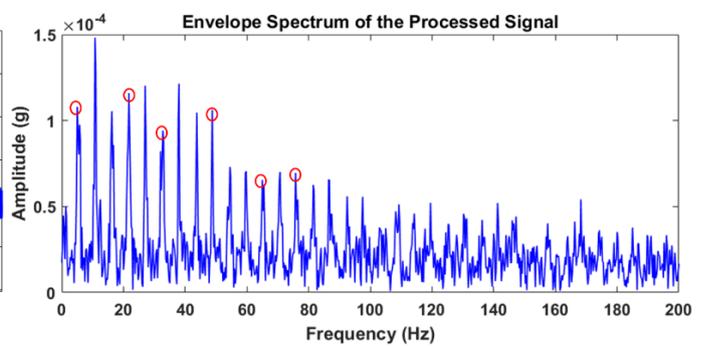

(c)

**Fig. 8.** Vibration signal from the defective bearing processed by fast kurtogram (a) kurtogram (b) Processed signal (c) Envelope spectrum of the processed signal

In the instance of a two-stage gearbox, the kurtogram-based approach has identified just one sudden event in the time waveform as depicted in Fig. 9. This incident represents an interference with significantly higher amplitude than the background noise. Consequently, it resulted in an increased kurtosis value and improved signal-to-noise ratio to some degree. The gear fault frequency is discernible in the envelope spectrum derived from the Kurtogram-based technique;

however, only two harmonics were detected using this method, which is considerably fewer compared to the recommended approach.

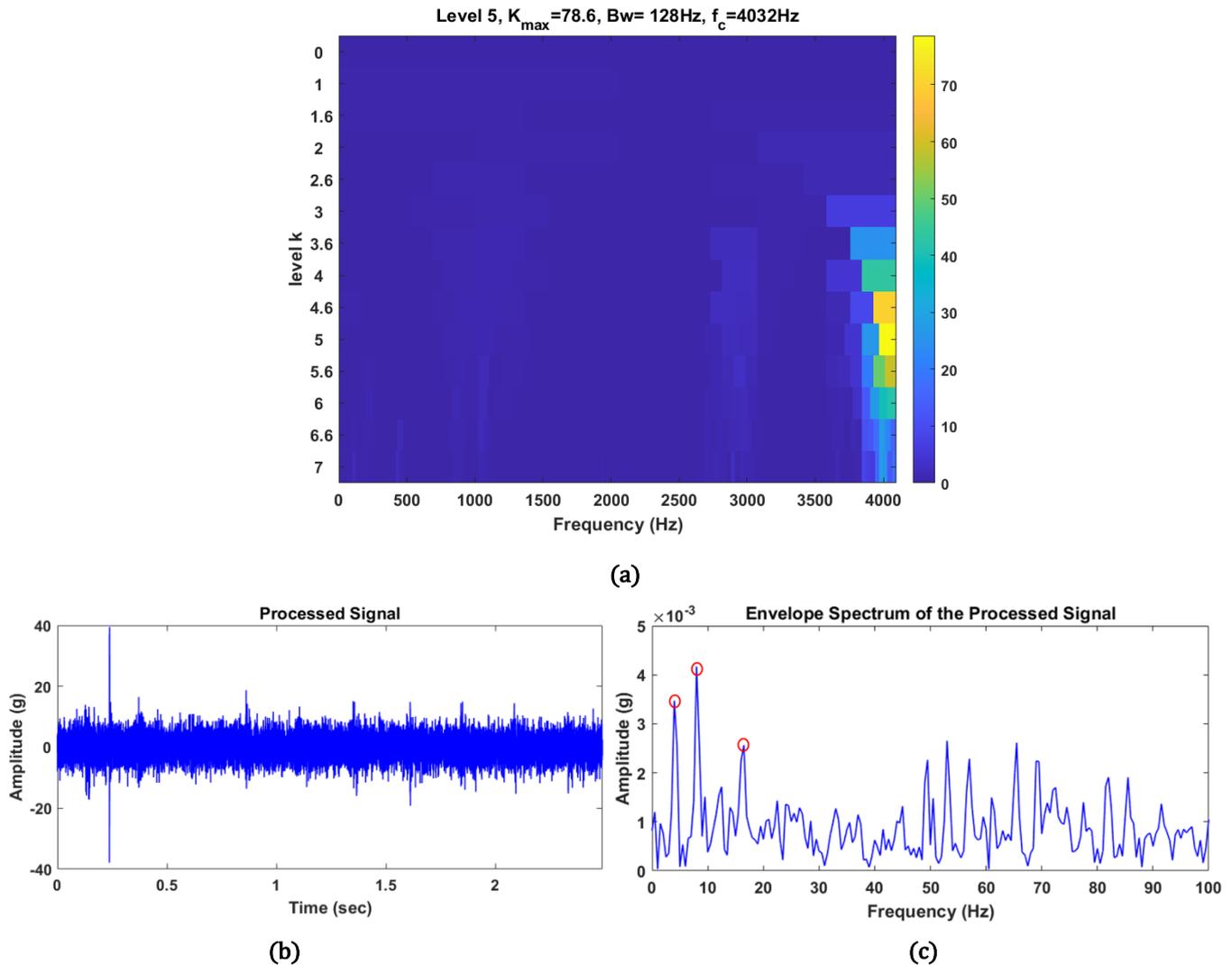

Fig. 9. Vibration signal from two-stage gearbox processed by fast kurtogram (a) kurtogram (b) Processed signal (c) Envelope spectrum of the processed signal

The fast-spectral coherence approach has also been considered for comparison. Both the signals from the faulty bearing and two-stage gearbox have undergone fast spectral coherence analysis. The resulting cyclic map and their envelope spectrum are depicted in Fig. 10 and Fig. 11 respectively. In the case of the defective bearing, periodic impulses are clearly visible in the cyclic map, along with the observation of fault frequency and its harmonics in the envelope spectrum. However, other frequencies besides fault frequency are also noticeable, potentially impacting the fault diagnostic process due to the lack of time domain waveform information in the fast spectral coherence method; hence an envelope spectrum-based indicator (ENVSI) [33] is employed for validation purposes. For a defective bearing on the pulley, ENVSI obtained by our proposed method is 0.48 while fast spectral coherence yields an ENVSI value of only 0.29

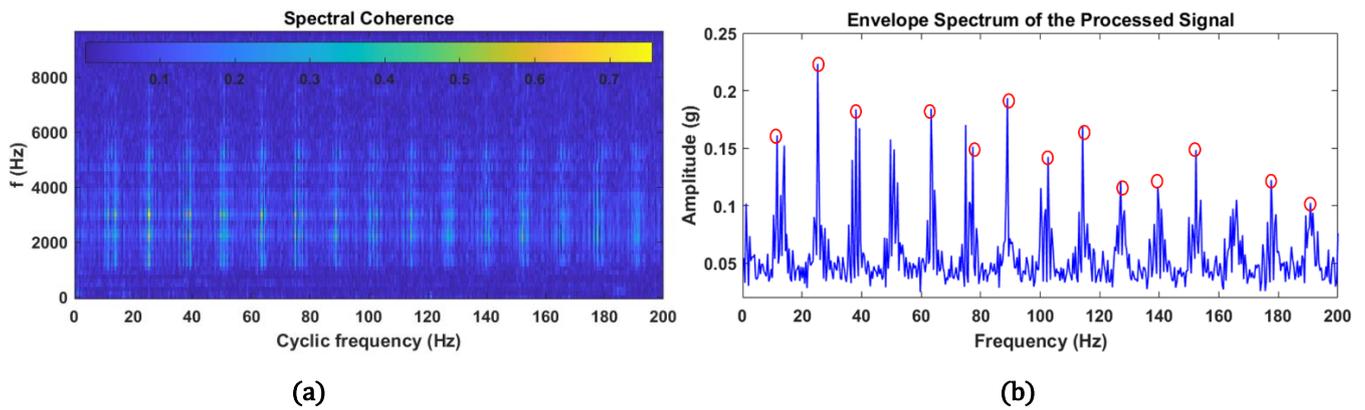

(a)                                               (b)

**Fig. 10.** Vibration signal from defective bearing processed by fast spectral coherence **(a)** spectral coherence map **(b)** Envelope spectrum of the processed signal

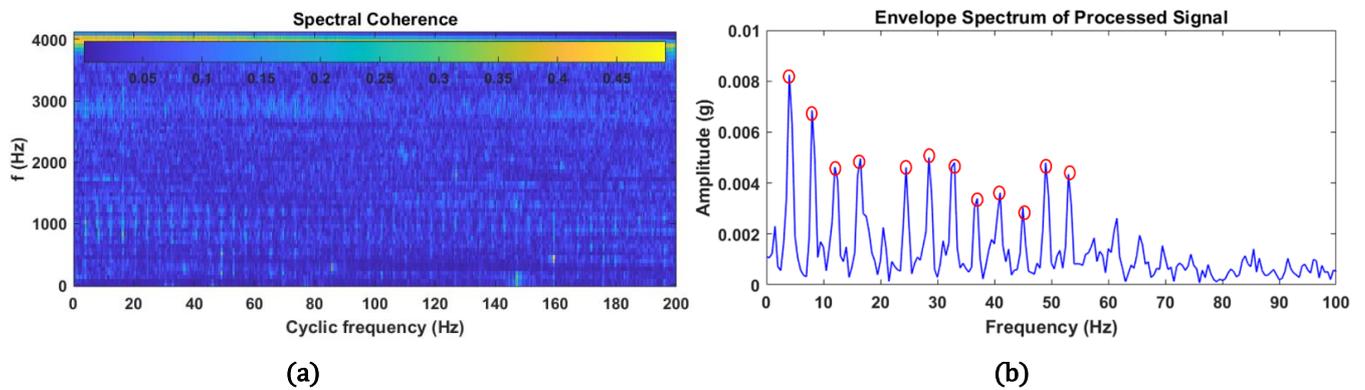

(a)                                               (b)

**Fig. 11.** Vibration signal from two-stage gearbox processed by fast spectral coherence **(a)** spectral coherence map **(b)** Envelope spectrum of the processed signal

In Fig. 11, the cyclic map and envelope spectrum for the two-stage gearbox are displayed. It is evident in Fig. 11 (a) that only a few impulses are visible in the cyclic map. The fault frequency and its harmonics can be easily identified in the spectrum. However, compared to the suggested approach, there are very few harmonics of the fault frequency. Therefore, it is concluded that the proposed method is more effective in capturing additional harmonics of the fault frequency to provide a clearer representation of the defect. Additionally, ENVSI is calculated to validate the effectiveness of this method where an ENVSI value of 0.56 is obtained as opposed to 0.32 with the fast spectral coherence method. The amplitude of frequencies also appears much smaller than those obtained by using our proposed scheme.

The results of the comparison validate the efficacy and usefulness of the suggested methodology for both the bearing and gearbox.

## 6. Conclusion

In this work, an optimized Morlet wavelet filter for the diagnosis of bearing defects in belt conveyor drives is put forward. The CAO is utilised to optimize the parameters of the wavelet filter considering correlated kurtosis (CK) as a fitness function. The optimized wavelet filter wavelet filter is better able to locate the informative frequency band to extract the repetitive transients. The efficacy and robustness of the proposed methodology have been tested on the signals obtained from the defective bearing and the two-stage gearbox of the belt conveyor system. The proposed methodology has also been compared with the existing methods including fast kurtogram and fast spectral coherence to validate the effectiveness of the proposed method. The results obtained suggested that the proposed methodology is an effective tool for detecting the informative frequencies in the presence of background noise and random interferences.

Acknowledgements

"The work is supported by the National Center of Science, Poland under Sheng2 project No. UMO-2021/40/Q/ST8/00024 NonGauMech - New methods of processing non-stationary signals (identification, segmentation, extraction, modeling) with non-Gaussian characteristics for the purpose of monitoring complex mechanical structures".